\documentclass[10pt, sigconf]{acmart}
\usepackage[]{todonotes} 
\usepackage[ruled, vlined, linesnumbered, noend]{algorithm2e}
\usepackage[super]{nth}
\usepackage{mathtools}
\usepackage{xcolor}
\usepackage{booktabs}
\usepackage{amsfonts}
\usepackage{hyperref}

\acmYear{2024}\copyrightyear{2024}
\setcopyright{acmlicensed}
\acmConference[EuroMLSys '24]{4th Workshop on Machine Learning and Systems}{April 22, 2024}{Athens, Greece}
\acmBooktitle{4th Workshop on Machine Learning and Systems (EuroMLSys '24), April 22, 2024, Athens, Greece}
\acmDOI{10.1145/3642970.3655833}
\acmISBN{979-8-4007-0541-0/24/04}

\def\@copyrightspace{\relax}

\newcommand{\namex}{Sponge\xspace}
\newcommand{\ballnumber}[1]{\tikz[baseline=(myanchor.base)] \node[circle,fill=.,inner sep=1.3pt] (myanchor) {\color{white}\bfseries\footnotesize #1};}

\title{\huge{\namex{}: Inference Serving with Dynamic SLOs Using In-Place Vertical Scaling}}

\author{Kamran Razavi}
\authornote{Equal Contribution}
\affiliation{%
 \institution{Technical University of Darmstadt}
}
\author{Saeid Ghafouri}
\authornotemark[1]
\affiliation{%
 \institution{Queen Mary University of London}
}
\author{Max M\"uhlh\"auser}
\affiliation{%
 \institution{Technical University of Darmstadt}
}
\author{Pooyan Jamshidi}
\affiliation{%
 \institution{University of South Carolina}
}
\author{Lin Wang}
\affiliation{%
 \institution{Paderborn University}
}



\begin{document}

\begin{abstract}

Mobile and IoT applications increasingly adopt deep learning inference to provide intelligence. Inference requests are typically sent to a cloud infrastructure over a wireless network that is highly variable, leading to the challenge of dynamic Service Level Objectives (SLOs) at the request level. 

This paper presents \namex{}, a novel deep learning inference serving system that maximizes resource efficiency while guaranteeing dynamic SLOs. \namex{} achieves its goal by applying in-place vertical scaling, dynamic batching, and request reordering. Specifically, we introduce an Integer Programming formulation to capture the resource allocation problem, providing a mathematical model of the relationship between latency, batch size, and resources. We demonstrate the potential of \namex{} through a prototype implementation and preliminary experiments and discuss future works.

\end{abstract}

\maketitle
\thispagestyle{plain}
\pagestyle{plain}

\section{Introduction}
Within the domain of mobile and IoT applications, cloud-based Deep Learning (DL) inference plays an important role, with user satisfaction and resource efficiency serving as key performance indicators~\cite{liu2019edge, kepuska2018next, 2020-nsdi-carmap}. Since most DL-powered applications involve user interaction, they must comply with strict requirements on the inference latency, a.k.a. meeting the Service Level Objectives (SLOs) of the inference request. On the other hand, the resources needed to provision such a DL inference serving system should be minimized to reduce the cost~\cite{2020-osdi-serving, razavi2022fa2, gujarati2017swayam, salmani2023reconciling, 2019-eurosys-grandslam, 2020-socc-inferline}.


SLOs are comprehensively defined from end to end, with the variable network time required for transferring user requests and input data introducing dynamic time budgets for serving inference requests. Therefore, when setting expectations for mobile and IoT applications, it is important to define SLOs that cover both the network and computing aspects from start to finish. Ignoring the time it takes for information to travel through the network, inference serving systems may find themselves with not enough time to handle requests properly, resulting in SLO violation. Hence, resource allocation must consider a variety of time budgets of a single user using the same application. Managing this dynamism poses a critical challenge for inference serving systems, where the effective handling of diverse SLOs and the consideration of fluctuating network conditions are imperative to ensure the fulfillment of end-to-end SLOs.

Existing inference serving systems mostly consider only the inference part with static SLOs, i.e., all requests have the same SLOs when they reach computing units. Their horizontal scaling-based approach cannot incorporate diverse SLOs at the request level~\cite{2020-socc-inferline, 2020-socc-gslice, 2020-osdi-serving}. For example, FA2~\cite{razavi2022fa2} adjusts the number of minimum-resource instances to achieve the highest resource efficiency (throughput). Moreover, bringing new instances in horizontal scaling ties with the cold-start issue (a few seconds~\cite{romero2019infaas, ghafouri2024ipa}), which cannot cope with the dynamically changing network conditions.
Jellyfish~\cite{vinod2022jellyfish}, on the other hand, aims to guarantee end-to-end SLOs while achieving high inference accuracy by using pre-loaded model-switching and trading accuracy for latency, which may not always be possible for all applications.

We propose a new system, \namex{}, aiming to address this research gap. Our main insight is that the combination of in-place vertical scaling, dynamic batching, and request reordering is a powerful tool to combat request-level dynamism. In particular, the new in-place vertical scaling feature of Kubernetes~\cite{kubernetes_in_place} allows developers to resize CPU/memory resources allocated to containers without restarting them, eliminating the cold-start issues of vertical scaling, while request reordering allows for requests with a lower remaining time budget to be processed earlier. At the same time, dynamic batching increases the system utilization to further reduce the needed computing resources. We formulate the problem and propose a method for inference serving with dynamic SLOs. \namex{} relies on three adaptation strategies to capture per-request dynamic SLOs: \ballnumber{1} in-place vertical scaling to change the computing resources of DL models in spot, \ballnumber{2} request reordering to prioritize close-to-deadline requests, and \ballnumber{3} dynamic batching to increase the utilization of the DL models. More specifically, \namex{} achieves dynamic SLOs guarantee and high resource utilization by first providing a mathematical relation between vertical scaling with batching and processing latency of the DL model using historical data and then designing a request-based mathematical modeling of the entire framework to guarantee SLOs of all requests while minimizing the resources. Furthermore, we propose a simple algorithm for small cases to iterate over all possible configurations and find optimal resource and batch size allocations. The preliminary experimental results show a reduction in over 15$\times$ of the SLO violation compared to the existing approaches.

\namex{} currently does not consider pipelines of DL models. Complex applications such as intelligent virtual assistants consist of multiple DL models, coordinated with a Directed Acyclic Graph (DAG), collaboratively generating a meaningful output. Such applications require a more intricate solution due to data dependencies among DL models, resulting in a strong coupling of scaling decisions for different DL models. Furthermore, vertical scaling sustains workloads to some extent due to the DL model parallelization level and availability of computing resources in a sine node. Therefore, multiple instances of the same DL model (horizontal scaling) may need to reside in different computing nodes to support the incoming workload. We consider these directions as future works of \namex{}.

This paper contributes by discussing the challenges of dynamic SLOs on DL inference serving systems. Then, we
\begin{itemize}
    \item present the design of \namex{}, a new DL inference serving system for dynamic SLOs based on the idea of in-place vertical scaling, request reordering, and dynamic batching.
    \item provide an Integer Programming formulation to encapsulate the problem of dynamic SLOs by introducing a mathematical modeling of the relation between latency, batch, and CPU in inference serving systems.
    \item build a prototype system for \namex{}~\footnote{\url{https://github.com/saeid93/sponge}} and evaluate it using 4G/LTE bandwidth logs datasets. \namex{} reduces the SLO violation by over $15\times$ compared to a horizontal state-of-the-art autoscaler.
\end{itemize}
\section{Motivation} \label{sec:motivation}
In this section, we first discuss the challenges raised by variable networks and then identify the challenges in efficient in-place vertical scaling.

\subsection{Dynamic SLO}
\begin{figure}
    \centering
    \includegraphics[width=0.48\textwidth]{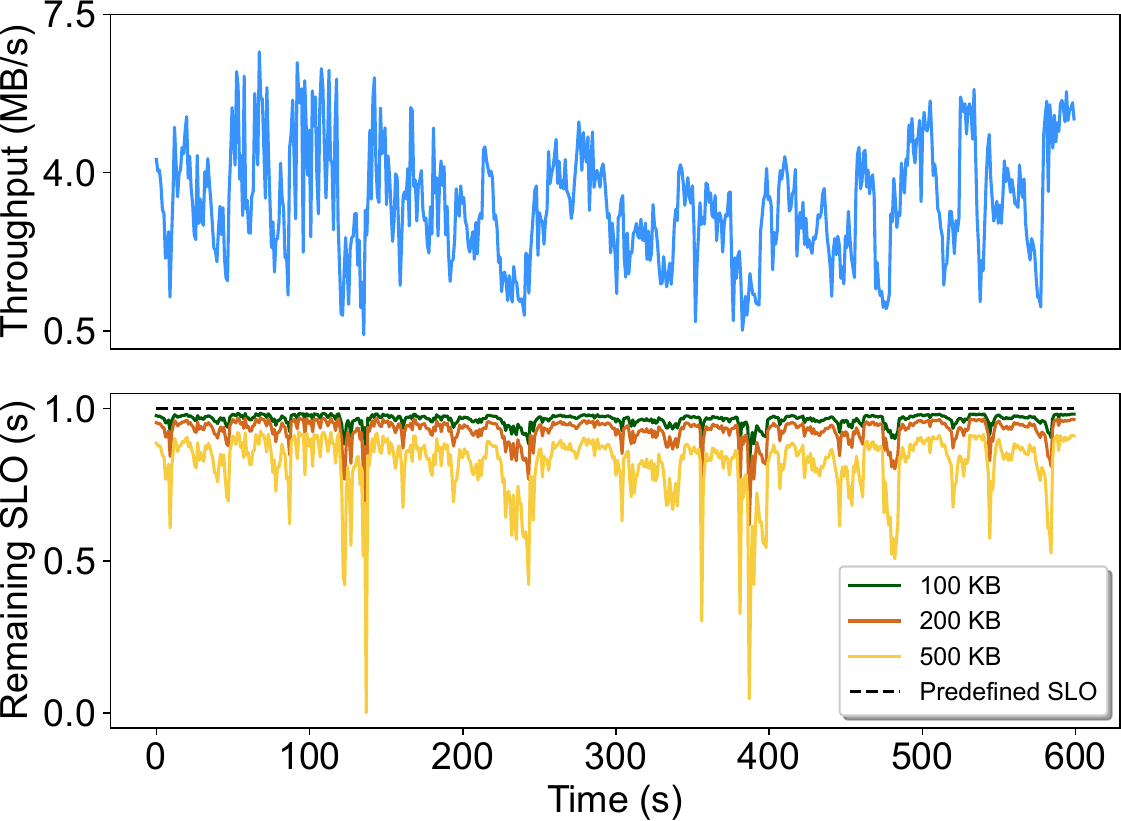}
    \caption{Bandwidth measurements in 4G networks provided by~\cite{vanderHooft2016}. The bandwidth varies from 0.5MB/s to ~7MB/s in a 10-minute range (top figure). The below figure demonstrates the remaining SLO for processing when the user sends a 100 KB, a 200 KB, or a 500 KB image over the same network's bandwidth.}\label{fig:dynamic_slo}
    \vspace{-0.4cm}
\end{figure}

Fluctuations in network bandwidths, e.g., caused by user mobility, are inevitable~\cite{ghafouri2022mobile, brandherm2022bigmec}, as illustrated in Figure~\ref{fig:dynamic_slo} (top). This variability influences the transmission overhead associated with sending data across the network for remote processing, leading to a reduction in the time budget available for server-side deployed services, as depicted in Figure~\ref{fig:dynamic_slo} (bottom). Consequently, service providers are compelled to account for network latency to ensure compliance with the end-to-end latency requirements specified in the SLO.

We use a simple human detection model trained on the ResNet architecture to motivate this work, where it detects a human in an image of 200 KB while the requests are being sent on a dynamically changing network (e.g., 4G) under a static workload of 100 requests per second with the SLO of 1000 ms (similar conditions to Figure~\ref{fig:dynamic_slo}). Table~\ref{tab:core} shows the execution latency of the model with different allocated CPU cores and batch sizes when considering the SLO. For calculating the required number of instances per instance type (different core numbers), we divide the incoming workload by the throughput of an instance. 
Following the approach in FA2~\cite{razavi2022fa2}, where they use one-core instances, we need five instances to process a batch of 2 requests per 97~ms, which means that we can process a batch of 10 requests, or 20 requests per second (RPS), over a 1000~ms SLO. This approach works perfectly if the network is static. However, if the network latency takes up to half of the SLO, FA2 will drop all the requests, as there is no possible solution in their approach with one-core instances, even with the smallest batch size. Furthermore, even if the network latency takes just 40~ms, the system needs to bring up a new instance to avoid dropping requests or violating the SLO, meaning that the system will suffer from the cold start of a new instance until the system stabilizes again. Alternatively, meeting the SLO in the context of a dynamically changing network bandwidth could have been achieved through the dynamic modification of computing resources within the instance (in-place vertical scaling). In the same scenario, if we had up to 600~ms of network delay, we could still serve the requests without violating or dropping any request by changing the instance core from 1 core to 8 cores with a batch size of 4. InfAdapter~\cite{salmani2023reconciling} employs profiling data to determine CPU core allocation for DL models. For instance, under a workload of 100 RPS, the model's computing resources and batch size remain static. However, when faced with changes in the SLO, it switches to a different model variant with predefined CPU core allocation, encountering similar challenges as FA2 (cold start and static CPU core allocation). 

\begin{table}[]
    \centering
    \caption{Execution latency (P99) of a ResNet model (human detector) with different CPU cores using different batch sizes while guaranteeing SLO of 1000~ms under the workload of 100 RPS. 
    }
    \small
    \vspace{-0.4cm}
    \begin{tabular}{@{}r r r r r@{}}
        \toprule
        Cores & Batch & Latency (ms) & Throughput (RPS) & Total Cores\\
        \midrule
         1 & 1 & 55 & $18 \times 6 = 108$ & $1 \times 6 = 6$\\
         1 & 2 & 97 & $20 \times 5 = 100$ & $1 \times 5 = 5$\\
         2 & 4 & 94 & $40 \times 3 = 120$ & $2 \times 3 = 6$\\
         4 & 8 & 92 & $80 \times 2 = 160$ & $2 \times 4 = 8$\\
         8 & 4 & 37 & $108 \times 1 = 108$ & $1 \times 8 = 8$\\
         8 & 8 & 62 & $128 \times 1 = 128$ & $1 \times 8 = 8$\\
         \bottomrule
    \end{tabular}
    \label{tab:core}
\end{table}

\subsection{Autoscaling Challenges} \label{sec:challenges}
Creating an effective in-place vertical scaling system for DL inference serving is a complex task. Precisely, we pinpoint the following challenges, which collectively differentiate the scaling problem in DL inference serving systems from those examined in other systems.

\noindent\textbf{Dynamic SLO at the request level.}
In wireless networks conditions can change over time. This can be due to various factors, such as changes in network traffic, hardware performance, signal strength, and resource availability~\cite{xu2020understanding, huang2012close}.
These factors can cause variable delays in network transmission for inference requests, leading to requests with dynamic SLOs. Accommodating dynamic SLOs at the request level requires fine-grained control over resource allocation to ensure each request meets its SLO. This level of granularity is challenging to achieve with vertical scaling since changing the resources to guarantee one request SLO affects all the requests' processing latency in the system.



\noindent\textbf{Batch size.}
DL inference serving systems commonly utilize request batching to enhance resource efficiency~\cite{2019-sosp-nexus,2017-nsdi-clipper,2020-arxiv-lazybatching, 2020-socc-inferline}. More precisely, batching can increase throughput as more tasks or requests can be processed in a given amount of time. Furthermore, batching can help meet latency constraints with dynamic batching policies, where batch sizes are determined online, during runtime, depending on the latency constraints of each application~\cite{razavi2022fa2, 2019-eurosys-grandslam}. However, it is important to note that large and small batch sizes can have drawbacks if not properly managed. Large batch sizes can critically violate the latency of many requests within a batch, while small batch sizes could cause excessive queuing and may not exploit potential opportunities for increased throughput.


In the next section, we provide an in-place vertical-based autoscaler to capture the discussed challenges by first discussing how to reconcile vertical scaling and batch size in the context of inference serving systems, and second, providing a mathematical formulation to mimic the autoscaling problem with the consideration of dynamic SLOs.

\section{System Design} \label{sec:design}
This section provides our solution for inference serving systems with dynamic SLOs. Our goal is to use minimal resources to provision the DL model with in-place vertical scaling, request reordering, and dynamic batch sizing while guaranteeing all the requests' SLO.

\subsection{Overview}
\namex{} consists of four components as is shown in Figure~\ref{fig:overview}:

\begin{figure}
    \centering
    \includegraphics[width=0.48\textwidth]{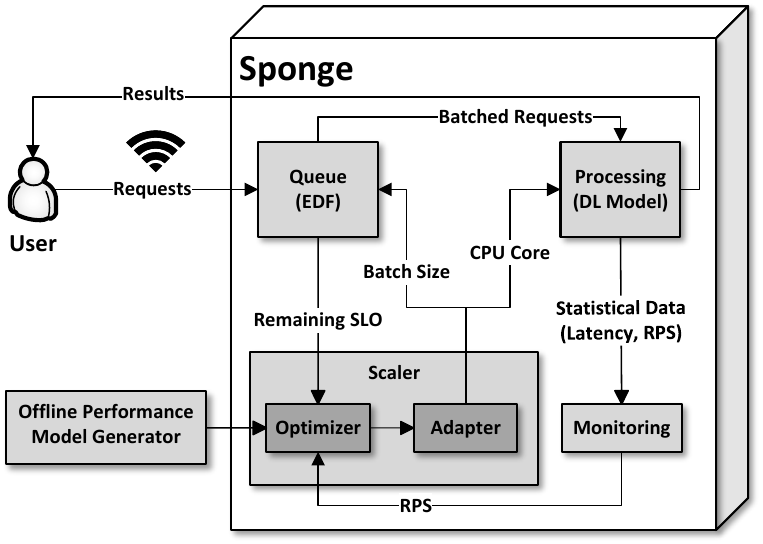}
    \caption{An overview of the \namex{} architecture. The monitoring service collects metric data from the DL model. The queue prioritizes requests according to the EDF policy. The scaler is responsible for determining vertical scaling and batch size decisions for the DL model and adjusting the system accordingly.}\label{fig:overview}
\end{figure}

\noindent\textbf{Monitoring.}
The monitoring component uses Prometheus~\cite{prometheus} to observe the incoming workload to the system. It will monitor the workload destined for the model on a predefined time interval. Additionally, it receives the end-to-end request latency from the processing component to calculate the SLO violation rate and the accuracy of the performance model.

\noindent\textbf{Queuing.}
The queuing component receives the request from the user, reorders the request based on the remaining SLO (Earliest Deadline First (EDF)), and creates a batch with the given batch size from the solver. In addition, it sends the set of requests with their communication latency to the optimizer.

\noindent\textbf{Processing.}
The processing component has the computing power to execute inferences. It receives batches from the queue, processes them, and sends them to the user. Furthermore, it sends the statistical data (queuing latency and processing latency) to the monitoring component.

\noindent\textbf{Scaler.}
The scaler component first aims to find the vertical scaling CPU cores and batch size decisions to achieve the highest resource efficiency while respecting all the request SLOs in the system by using the workload (reported by the monitoring component) and the remaining SLOs of all the requests after being reordered by the queuing component in the optimizer. Next, its adapter part adjusts the system by sending a signal to the processing component with the new CPU core allocation and a signal to the queueing component with the new batch size configuration.

\subsection{Performance Model} \label{sec:performance}
For effective decision-making within the solver, \namex{} needs knowledge of the performance metrics, specifically the throughput $h(b, c)$ and latency $d(b, c)$, associated with the DL model. Previous research has indicated that the performance of DL inference tends to be highly predictable~\cite{2019-eurosys-grandslam,2017-nsdi-clipper,2018-icdcs-rightsizing, 2020-osdi-serving}.
We follow the same line and use profiling data and robust regressions~\cite{1981-cacm-regression} to build a model for any given DL model.
GrandSLAm~\cite{2019-eurosys-grandslam,2017-nsdi-clipper} suggests a linear relationship between batch size and latency, that is, $l(b, c) = \alpha_1 \times b + \beta_1$, and FA2~\cite{razavi2022fa2} suggests a second-order quadratic polynomial for a lower total MSE. However, none of the above works consider changes in the computational resources (e.g., number of CPU cores) of the DL models. For simplicity, we use the linear relation in the current work.

To have a relation between latency and CPU, we use Amdahl's law~\cite{amdahl1967validity} for latency prediction under a given batch size:
\vspace{-5pt}
\begin{equation} \label{eq:latency_core}
    L(b, c) = \frac{\alpha_2}{c} + \beta_2
\end{equation}

Equation~\ref{eq:latency_core} states an inverse relation between the number of CPU cores and latency if the model can use additional CPU cores, which is the case in ML models.

On the other hand, the linear relation of batch size and latency suggests that $\alpha_1$ and $\beta_1$ have inverse relations with CPU cores, e.g., $\alpha_1 = \gamma_1/c + \delta_1$ and $\beta_1 = \epsilon_1/c + \eta_1$ (otherwise, $l(b, c)$ would become linear in Equation~\ref{eq:latency_core}). Therefore, to incorporate computational resources into batch/latency profiling, we combine the linear relation of batch/latency and the inverse relation of CPU/latency as follows.

\vspace{-5pt}
\begin{equation}
    \begin{split}
    \label{eq:latency_batch_core}
        l(b, c) = & (\frac{\gamma_1}{c} + \delta_1) \times b + \frac{\epsilon_1}{c} + \eta_1 \\
        = & \frac{\gamma_1 \times b}{c} + \frac{\epsilon_1}{c} + \delta_1 \times b + \eta_1
    \end{split}
\end{equation}

\begin{figure}
    \centering
    \includegraphics[width=0.48\textwidth]{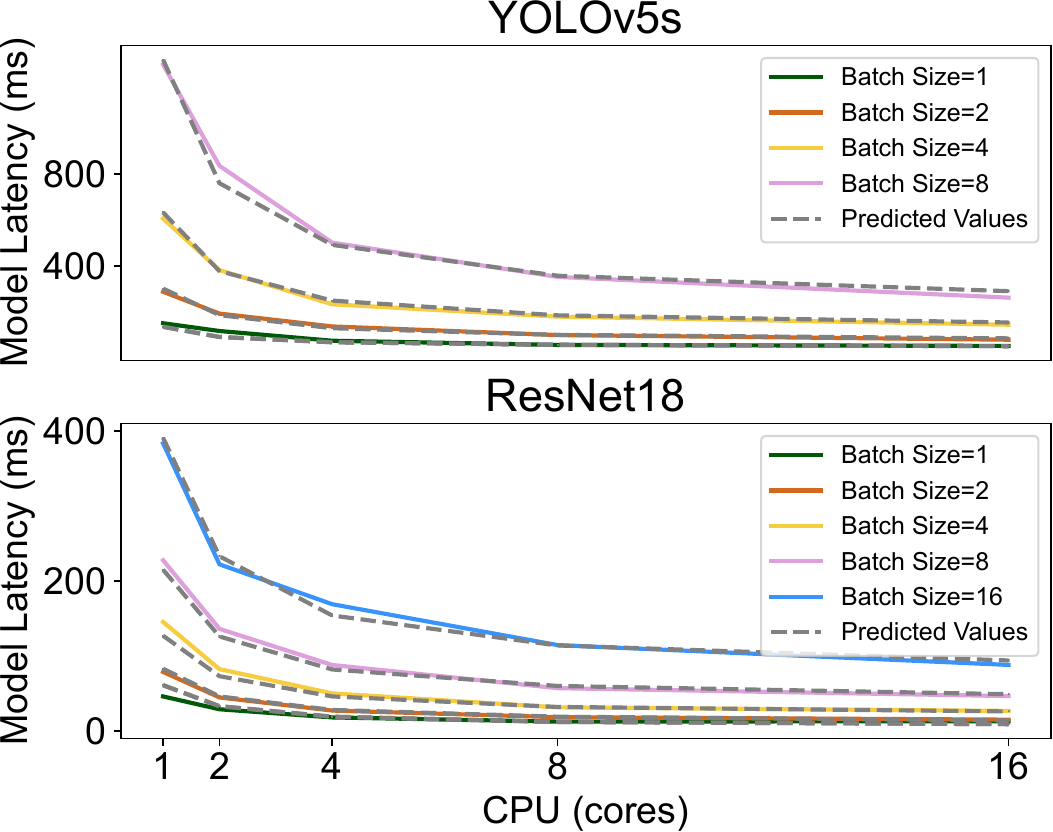}
    \caption{Latency vs. different CPU core allocations and batch sizes using real and predicted for the YOLOv5n and ResNet18 DL models.}\label{fig:vertical_batch_latency}
    \vspace{-0.4cm}
\end{figure}

Our preliminary evaluation with the data sets profiled from ResNet18 and YOLOv5n models used in Figure~\ref{fig:vertical_batch_latency} confirms that the latency/CPU/batch model in Equation~\ref{eq:latency_batch_core} provides a realistic estimation of latency with different CPU cores and batch sizes on different DL models. The throughput of a DL model is directly given as a function of batch size and CPU cores, e.g., $h(b, c) = b/l(b, c)$.

\subsection{Problem Formulation}
\begin{table}[!t]
    \centering
    \footnotesize
    \caption{Notations}\label{table:notation}
    \begin{tabular}{ p{1cm}p{6cm}  }
        \toprule
        \textbf{Symbol} & \textbf{Description}\\
        \midrule
        $R$ & Set of all requests \\
        $b$ & Model's batch size \\
        $c$ & Model's CPU allocation \\
        $cl_r$ & Communication latency associated with $r \in R$ \\
        $cl_{max}$ & Highest $cl_r$ in $R$ \\
        $SLO$ & Pre-defined SLO for $R$ \\
        $l(b, c)$ & Processing time of a model with allocation core $c$ and batch size $b$ \\
        $q_r(b, c)$ & Queuing time of $r \in R$ with allocation core $c$ and batch size $b$\\
        $h(b, c)$ & Throughput of a model with allocation core $c$ and batch size $b$ \\
        $\lambda$ & Request arrival rate \\
        \bottomrule
    \end{tabular}
\end{table}
The optimizer generates scaling decisions by solving an optimization problem.
Now, we provide a formal formulation for the problem given that the end-to-end latency for a request is the aggregation of the communication latency (the time the request takes to be received by the system from the user device), the queuing (the time the request spends in the queue before being processed), and the processing latencies (inference latency) of the request.

Suppose that we are given a model and a set of requests $R$ with a predefined SLO. Each request $r \in R$ has communication latency $cl_r$. 
The arrival rate of the application request is denoted by $\lambda$.
Due to the instability of the network, as we have already discussed in Section~\ref{sec:motivation}, we apply the earliest-deadline-first (EDF) queue ($q(b,c)$), similar to GradnSLAm~\cite{2019-eurosys-grandslam}, since request reordering prioritizes the processing of requests with lower remaining SLOs due to their more stringent completion deadlines.

Let us denote the number of CPU cores allocated and the batch size of the model by $c$ and $b$, respectively. In addition, we use $cl_{max} = max(cl_r, r \in R)$ to indicate the highest communication latency in the current requests.

The monitoring system continuously reports to the adapter the average number of requests served by the model in a given period.
To ensure the stability of the system, that is, no back pressure should form in the queue, and the throughput of the model should be no less than the expected request rate, that is, $h(b, c)  \geq \lambda$.
Such a constraint ensures that the model is sufficiently provisioned.
As a result, the queuing of requests on the model will be under control.

The optimization problem is to decide $c$ and $b$ for the model such that under the workload $\lambda$, none of the request SLOs are violated. 
The goal is to minimize the amount of resources (CPU cores) used for the model. The problem can be formulated with the following integer program (IP):
\vspace{1pt}
\begin{equation}
    \begin{aligned}
        \text{Minimize} \quad & c + \delta \times b\\
        \text{subject to} \quad & l(b, c) + q_r(b, c) + \text{cl}_{max} \leq
        SLO, \quad \forall r \in R \\
        & h(b, c) \geq \lambda \\
        & b, c \in \mathbb{Z}^+ \\
    \end{aligned}
\end{equation}

In the objective function, in addition to CPU cores, we incorporate an insignificant penalty term $\delta$ into the batch size to mitigate unnecessary latencies.
The first constraint ensures that all requests for SLOs, including communication latency, will be satisfied. We use the smallest SLO in the current batch for all requests in the same batch because we do not intend to violate any remaining SLO requests.
The second and third constraints are designed to maintain system stability, necessitating that the CPU cores and the batch size be constrained to positive integer values.
The objective is to minimize the total amount of resources, that is, the total number of CPU cores given to the model. 
All the notation used is available in Table~\ref{table:notation}.

\subsection{Solution}
With IP and a single model, we use a brute force approach shown in Algorithm~\ref{algo:dp}. We feed the requests with their remaining SLOs to a queue and then reorder them based on the EDF policy (lines 1--2). After finding the maximum communication latency in the set of requests (line 4), we then iterate over all possible batch sizes and CPU core allocations (lines 5--6). Furthermore, we check if the current configuration and all the requests in the subsequent batches will satisfy their remaining SLOs (lines 10--15). Note that there will be a waiting time for the subsequent batches equal to the processing latency of the previous batches, calculated in line 14. Finally, if there is no objection against the current batch size and CPU core allocation configurations (line 15), we send the found configuration to be enforced to the system. The algorithm generates the optimal CPU core allocation with the smaller batch size with the current allocation, since it iterates from 1 to the maximum CPU core and batch size allocations. 

\begin{algorithm}[!t]
\small
\SetAlgoLined
\SetKwInOut{Input}{input}\SetKwInOut{Output}{output}
\SetKwIF{If}{ElseIf}{Else}{if}{then}{else if}{else}{end if}%
\Input{SLO, Set of requests $r \in R$ with communication latency, Performance model} 
\Output{$c, b$}
$q \leftarrow R$\\
Reorder q (EDF policy)\\
$n = len(R)$\\
Calculate $cl_{max}$\\
\For{$c$ in $[1,c_{max}]$}{
    \For{$b$ in $[1,b_{max}]$}{
        Calculate $l(b, c)$\\
        $better=True$\\
        $q_r = 0$\\
        \For{$i$ in $[1,n,b]$}{
            \uIf{$l(b, c) + cl_{max} + q_r \geq SLO$}{
                $better=False$\\
                \textbf{break}
            }
            $q_r = q_r + l(b, c)$\\

        }
        \uIf{$better=True$}{
                \textbf{return} c, b
            }
    }
}
 \caption{Optimal CPU and batch size finder}
 \label{algo:dp}
\end{algorithm}

\section{Preliminary Evaluation} \label{sec:evaluation}

\namex{} is implemented in 6K lines of Python. For evaluation, we use a physical machine from Chameleon Cloud~\cite{keahey2020lessons} equipped with Intel(R) Xeon(R) Gold 6240R (48 threads). To enable the in-place vertical scaling, we install the experimental branch of minikube~\cite{minikube} since the in-place vertical scaling feature is not yet in the official releases~\cite{kubernetes_in_place}.


\noindent\textbf{Baseline.}
We compare \namex{} with a state-of-the-art horizontal autoscaler in inference serving systems, FA2, and static 8-core and 16-core instances. All approaches (including \namex{}) use a YOLOv5s~\cite{yolov5} with the performance modeling in Figure~\ref{fig:vertical_batch_latency} to detect humans in images. We also set $b_{max}$ and $c_{max}$ to $16$ for \namex{} as there is no significant gain afterward. For the adaptation period, we set one second same as the network bandwidth interval in the dataset.

\noindent\textbf{Workload generator.} In order to assess \namex{} in scenarios with dynamic network bandwidth, we design a workload generator that produces requests asynchronously at a fixed rate of 20 RPS with predefined SLOs similar to Figure~\ref{fig:dynamic_slo}. We use gRPC~\cite{grpc} to handle communication between all components of the system, including the workload generator.

\begin{figure}
    \centering
    \includegraphics[width=0.48\textwidth]{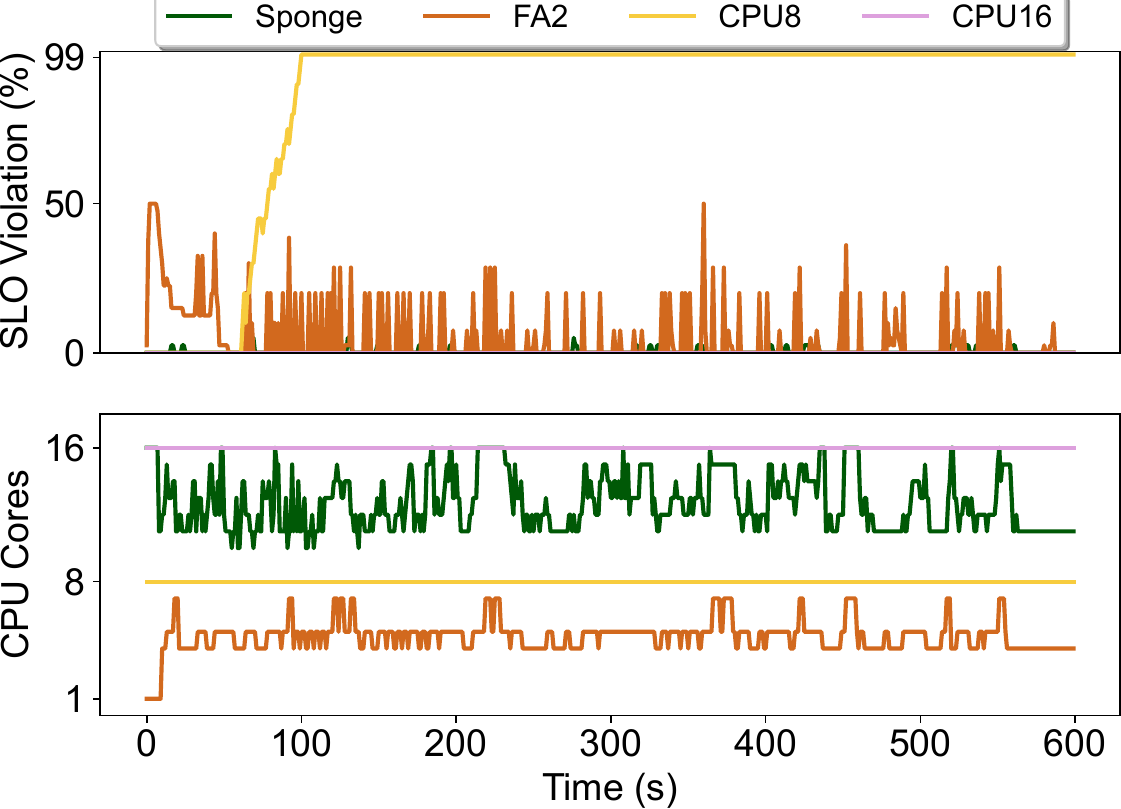}
    \caption{SLO violations and allocated CPU cores.}\label{fig:preliminary_evaluation}
    \vspace{-0.5cm}
\end{figure}

\noindent\textbf{Performance evaluation.} Figure~\ref{fig:preliminary_evaluation} demonstrates the overall performance of \namex{}, FA2, and statically assigned CPU cores under a dynamic network bandwidth. Under a given workload and the remaining SLOs, FA2 violates a large number of requests' SLO (roughly 5\% and over 50\% violation in some severe cases (Time = 1 and 360 in the same Figure) when the bandwidth becomes limited since bringing new instances is tied with the cold startup issue, and FA2 needs roughly 10 seconds to find a new configuration, adjust itself, and stabilize the system. The statically assigned 8-core instance experiences SLO violations after a few seconds due to insufficient computational resources to handle the requests, necessitating a more powerful instance. Conversely, the 16-core instance shows almost no SLO violations, indicating potential over-provisioning of the DL model. \namex{} solves the resource waste by dynamically changing the allocated CPU cores in response to the network bandwidth changes and reduces the amount of allocated CPU by over 20\% while sacrificing less that 0.3\% of SLO violations, compared to statically assigned 16-core instance.

\section{Related Work} \label{sec:related_work}
\noindent\textbf{Inference serving with SLO guarantee.}
Multiple works have been proposed with SLO guarantees~\cite{2020-osdi-serving, salmani2023reconciling, 2020-socc-gslice, 2019-sosp-nexus}. Model switch~\cite{zhang2020model} switches to a different model architecture in response to workload changes to ensure SLO. GrandSLAm~\cite{2019-eurosys-grandslam} uses dynamic batching and request reordering to increase system throughput with the SLO guarantee. InFaaS~\cite{romero2019infaas} gets user preferences about accuracy, cost, or performance and provides a model variant to satisfy the requested SLO. Jellyfish~\cite{vinod2022jellyfish} trades accuracy with latency by model switching and data adaptation to match the input of the model variant to guarantee latency SLO.

\noindent\textbf{Autoscaling in inference serving.} Autoscaling in inference serving has been extensively studied~\cite{hu2021scrooge, zhang2019mark, romero2021llama, 2020-socc-inferline}.
Kubernetes VPA~\cite{kubernetes_vpa} and HPA~\cite{kubernetes_hpa} use threshold-based metrics such as CPU or memory usage to change computing resources or the number of instances of DL-based inference services.
Clipper~\cite{2017-nsdi-clipper} provides an abstraction layer to simplify model deployment across frameworks and uses adaptive batching to increase system throughput.
IPA~\cite{ghafouri2024ipa} uses model switching and horizontal scaling to increase system accuracy while minimizing computing resources.
Cocktail~\cite{gunasekaran2022cocktail} uses a subset of model variants with a weighted scaling policy to ensure low cost, a predefined accuracy, and latency SLOs archived.
FA2~\cite{razavi2022fa2} uses graph transformation and dynamic programming to design a new horizontal autoscaler to increase system utilization with SLO guarantees.

The mentioned approaches neither consider dynamic networks (wireless and 4G/5G) without changing the model variant that affects other metrics such as cost and accuracy nor use in-place vertical scaling, which \namex{} has shown a necessity for state-of-the-art autoscalers to guarantee predefined latency SLO under a dynamic network bandwidth.
\vspace{-0.2cm}
\section{Conclusion \& Future Work} \label{sec:conclusion}
In this work, we presented \namex{}, the first inference serving system that uses in-place vertical scaling, request reordering, and dynamic batching with SLO guarantees. The preliminary evaluation shows that \namex{} reduces the SLO violation to 0.3\% while minimizing the CPU allocation in a dynamic network. We identify the following limitations of \namex{} and consider them as future directions:

\noindent\textbf{Model variant.} There are variations of the same DL model with different configurations in terms of architecture that are capable of doing similar tasks with different objectives such as accuracy~\cite{zhang2020model, romero2019infaas, vinod2022jellyfish}. Incorporating model variants requires careful system design, since the three pillars of accuracy, latency, and CPU allocation (even without vertical scaling) have conflicting relations~\cite{salmani2023reconciling}.

\noindent\textbf{Pipeline.} Many modern applications are composed of multiple DL models, such as Amazon Alexa, and are usually arranged as a DAG. Generalizing \namex{} to support such applications requires a new algorithm design, since there is a data dependency~\cite{razavi2022fa2, 2020-socc-inferline, hu2021rim, ghafouri2024ipa} between DL models and finding an optimal resource allocation for individual DL models requires consideration of all models in the system. 

\noindent\textbf{Multidimensional scaling.} The resource requirements of a DL model can be influenced by the dynamic nature of workloads~\cite{gujarati2017swayam, zhang2019mark}, making them difficult to predict. Vertical scaling can support the incoming workload to a certain degree, meaning that horizontal scaling must be considered if the workload is too much for a single instance of a DL model. The joint optimization of horizontal scaling and vertical scaling mechanisms brings new challenges, such as changing an upstream DL model's processing latency rate (vertical scaling), which affects the input rates on downstream DL models and may require additional instances (horizontal scaling). 

\section*{Acknowledgements} This work has been supported in part by NSF (Awards 2233873, 2007202, 2038080, and 2107463), Deutsche Forschungsgemeinschaft (DFG, German Research Foundation) – Project-ID 210487104 - SFB 1053, Roblox Corporation, and Chameleon Cloud.
\bibliographystyle{ACM-Reference-Format}
\bibliography{ref}

\end{document}